\documentclass[a4paper,11pt]{article}
\usepackage[utf8]{inputenc}
\pdfoutput=1
\usepackage{graphicx}
\usepackage{amsmath,amssymb,mathrsfs}
\usepackage{bbm}
\usepackage{color}
\usepackage{xcolor}
\usepackage{dsfont} 
\usepackage{cancel}
\usepackage{dsfont}
\usepackage{epstopdf}
\usepackage{epsfig}
\usepackage{bm}
\usepackage{authblk}
\usepackage{dcolumn}
\usepackage{enumitem}
\usepackage{multirow}
\usepackage{lineno}
\usepackage{mathtools}
\usepackage{url}
\usepackage{lineno}
\usepackage{wrapfig}
\usepackage{caption}
\usepackage{subcaption}
\usepackage{jheppub}
\usepackage[toc,page]{appendix}
\usepackage{wrapfig}

\makeatletter
\gdef\@fpheader{}
\g@addto@macro\bfseries{\boldmath}
\makeatother

\title{Bubble-wall velocity in local thermal equilibrium: hydrodynamical simulations vs analytical treatment}
\author[1]{Tomasz~Krajewski}
\author[2]{Marek~Lewicki}
\author[2]{Mateusz~Zych}
\affiliation[1]{Nicolaus Copernicus Astronomical Center, ul. Bartycka 18, 00-716 Warsaw, Poland}
\affiliation[2]{Faculty of Physics, University of Warsaw, ul.\ Pasteura 5, 02-093 Warsaw, Poland}

\emailAdd{tkrajewski@camk.edu.pl}
\emailAdd{marek.lewicki@fuw.edu.pl}
\emailAdd{mateusz.zych@fuw.edu.pl}
\date{February 2024}
\abstract{
We perform real-time hydrodynamical simulations of the growth of bubbles formed during cosmological first-order phase transitions under the assumption of local thermal equilibrium. We confirm that pure hydrodynamic backreaction can lead to steady-state expansion and that bubble-wall velocity in such case agrees very well with the analytical estimates. However, this is not the generic outcome. Instead, it is much more common to observe runaways, as the early-stage dynamics right after the nucleation allow the bubble walls to achieve supersonic velocities before the heated fluid shell in front of the bubble is formed. This effect is not captured by other methods of calculation of the bubble-wall velocity which assume stationary solutions to exist at all times and would have a crucial impact on the possible generation of both baryon asymmetry and gravitational wave signals. 
}

\begin{document}

\maketitle
\section{Introduction}
Phase transitions are present in a variety of particle-physics models. If they are first order they could create an environment for the generation of baryon asymmetry~\cite{Kuzmin:1985mm, Cohen:1993nk, Rubakov:1996vz, Morrissey:2012db} and production of a stochastic gravitational wave background, which could be potentially observed with the next generation of detectors ~\cite{Caprini:2015zlo, Caprini:2019egz, AEDGE:2019nxb, Badurina:2021rgt, LISACosmologyWorkingGroup:2022jok, Colpi:2024xhw}. Cosmological phase transitions are typically induced by quantum tunnelling or thermal fluctuation of a~scalar field between non-degenerate minima of its potential and proceed via nucleation of bubbles of the energetically favourable phase. Once the bubbles are nucleated, they start to grow, driven by the potential difference between the different phases inside and outside the bubble. On the other hand, particles crossing the bubble wall exert velocity-dependent friction on it. These two forces may balance each other, leading to steady-state expansion with constant velocity. %Otherwise, the bubble wall accelerates up to the moment of collision with other bubbles, which is often called a runaway scenario.
Otherwise, the bubble wall will continue to accelerate and reach a velocity very close to the speed of light\footnote{In local thermal equilibrium the acceleration will only stop upon collision with other bubbles although in realistic scenarios it is expected that out-of-equilibrium corrections can stop the acceleration first~\cite{Bodeker:2017cim, Azatov:2020ufh, Gouttenoire:2021kjv, Hoche:2020ysm}.}
Correct distinction between these two situations and accurate estimation of the bubble-wall velocity is crucial for both predictions of baryon asymmetry production in theories of Electroweak Baryogenesis %~\cite{Azatov:2021irb,Baldes:2021vyz}
and modelling of gravitational wave spectra produced in the transitions.

Backreaction coming from the particles interacting with the bubble wall was estimated by solving the Boltzmann equation for all the species in the plasma together with the equation of motion for the scalar field~\cite{Moore:1995ua, Moore:1995si, Dorsch:2023tss, Cline:2020jre, Laurent:2020gpg, Dorsch:2021ubz, Dorsch:2021nje, Cline:2021iff, Cline:2021dkf, Lewicki:2021pgr, Laurent:2022jrs, Ellis:2022lft}.
However, this method is quite challenging and the non-equilibrium part of the friction in the equation of motion is instead often parameterized with an effective phenomenological coefficient treated as a free parameter in hydrodynamical simulations~\cite{Kurki-Suonio:1995yaf, Hindmarsh:2013xza}. Recently, it has been shown that purely equilibrium hydrodynamic backreaction can inhibit the accelerating expansion~\cite{BarrosoMancha:2020fay,Ai:2021kak,Ai:2023see} and a simple estimate that can be interpreted as the upper limit on bubble-wall velocity for a given model has been derived. However, this approach, similarly to methods based on the Boltzmann equation, assumes that the plasma profiles are at any time fully developed into steady-state solutions with a given wall velocity~\cite{Espinosa:2010hh}. The mechanism preventing the further acceleration of the wall is connected with the heating of the plasma outside of the bubble and relies on this assumption heavily. 
In this work, we verify this assumption using hydrodynamic simulations without non-equilibrium friction.

As a benchmark model, we use a singlet scalar extension of the Standard Model, which is a simple and well-known example of a theory in which electroweak symmetry breaking may proceed as a~first-order phase transition~\cite{McDonald:1993ey, Espinosa:1993bs, Espinosa:2007qk, Profumo:2007wc, Espinosa:2011ax, Barger:2011vm, Cline:2012hg, Alanne:2014bra, Curtin:2014jma, Vaskonen:2016yiu, Kurup:2017dzf, Beniwal:2017eik, Beniwal:2018hyi, Niemi:2021qvp}. 
We use the temperature-dependent potential to find the solution corresponding to the nucleating bubble and simulate its evolution to observe as it develops into a self-similar solution.
We compare the results with the bubble-wall velocities and plasma profiles obtained with analytical methods. We have found that if a steady state expanding slower than the Jouguet velocity is reached, the two methods agree with very good accuracy. Such scenarios are, however, rare and fine-tuned. In the absence of non-equilibrium friction typical bubble accelerates beyond the Jouguet velocity before heating the plasma as the self-similar profile would suggest. As a result, most bubbles develop into very fast detonations which in the absence of non-equilibrium friction continue to accelerate corresponding to the so-called runaway solution. Thus, we find that velocity estimates assuming steady-state evolution may not be valid in the majority of cases and predictions of baryon asymmetry and gravitational wave emission based on them would be incorrect.

The paper is organised as follows. In section \ref{sec:tran_parameters} basic parameters describing cosmological phase transitions are defined, while in section \ref{sec:expansion} steady-state dynamics of the bubble expansion presenting different expansion modes are discussed. We review recent estimates of velocity \cite{Ai:2023see}, to which we refer further. In Section \ref{sec:singlet} we introduce the singlet scalar extension of the Standard Model and compute a high-temperature approximation of the effective potential of this theory. Then, we perform a scan of the parameter space determining transition parameters for different realisations of that model and estimate bubble-wall velocity from the matching equations. Section \ref{sec:dynamics} is devoted to the derivation of equations of motion adjusted to the scalar singlet model used in real-time simulations. Finally, in section \ref{sec:simulations} we show the results of the real-time simulations and compare them with the estimations mentioned above. We summarize our results in section \ref{sec:conclusions}. Details of our code and numerical methods used in the simulations are reviewed in the Appendix \ref{appendix}.

%%%%%%%%%%%%%%%%%%%%%%%%%%%%%%%%%%%%%%%%%%%%%%%%%%%%%%%%%%%%%%%%%%%%%%%%
\section{Transition parameters}\label{sec:tran_parameters}
%%%%%%%%%%%%%%%%%%%%%%%%%%%%%%%%%%%%%%%%%%%%%%%%%%%%%%%%%%%%%%%%%%%%%%%%
Cosmological first-order phase transitions proceed via nucleation
of bubbles of the broken phase from the background state of the symmetric, metastable phase. The probability of tunnelling per unit time and volume at temperature $T$ is given with the bubble nucleation rate~\cite{Coleman:1977py,Callan:1977pt,Linde:1980tt,Linde:1981zj}
\begin{equation}
\Gamma(T) = A(T)\textrm{e}^{-S}\, ,
\end{equation}
where for finite temperatures the Euclidean action $S=\frac{S_3}{T}$ and 
$A(T)=T^4\left(\frac{S_3}{2\pi T}\right)^{\frac{3}{2}}$. Nucleation temperature is defined as the value such that the probability of a true vacuum bubble forming within a horizon radius
grows close to unity~\cite{Ellis:2018mja}, i.e. 
\begin{equation}
N(T_n) = \int_{T_n}^{T_c} \frac{dT}{T} \frac{\Gamma(T)}{H(T)^4} \approx 1, 
\label{eq:nucleation}
\end{equation}
where $T_c$ denotes the critical temperature in which both minima are equally deep. Assuming that the transition is fast, one can neglect the change of the expansion rate of the Universe (assume that Hubble parameter $H(t)\approx$ const). Then, eq. \eqref{eq:nucleation} reduces to
\begin{equation}
\frac{S_3}{T_n}\approx 4\log\left(\frac{T_n}{H}\right),
\end{equation}
which for temperatures close to the electroweak scale gives $S_3/T_n \approx 140$~\cite{Caprini:2019egz}. To quantify the level of supercooling of the phase transition, we introduce the temperature ratio $T_n/T_c$. 

An important parameter describing the strength of the transition is the amount of latent heat released during the transition, typically normalised to the energy density of the cosmological background, resulting in \cite{Hindmarsh:2017gnf, LISACosmologyWorkingGroup:2022jok} 
\begin{equation}
    \alpha = \frac{1}{\rho_r}\left(\Delta V - \frac{T}{4}\Delta\frac{\partial V}{\partial T}\right) ,
\end{equation}
where $\Delta$ denotes the difference between symmetric and broken phases, and $V$ is temperature-dependent potential. For model-independent studies, one needs to define a generalized transition strength based on the trace of the energy-momentum tensor $\theta$~\cite{Giese:2020rtr}
\begin{equation}
    \alpha_{\theta} = \frac{\Delta\theta}{3 w_s},  \qquad\textrm{with}\qquad \theta = e - 3p
\end{equation}
where $w_s$ denotes enthalphy in the symmetric phase while $e$ and $p$ correspond to energy and pressure. Taking into account that the speed of sound might be different from the standard $c_s=1/\sqrt{3}$ and assuming only weak dependence on $T$, one can introduce pseudotrace $\bar{\theta}$ \cite{Giese:2020znk}. We then have a more accurate prescription for the transition strength
\begin{equation}
    \alpha_{\bar{\theta}} = \frac{\Delta\bar{\theta}}{3 w_s}\, , \qquad\textrm{with}\qquad \bar{\theta} = e -\frac{p}{c_b^2}\, ,
\end{equation}
where $c_b$ is the speed of sound in the broken phase.

%%%%%%%%%%%%%%%%%%%%%%%%%%%%%%%%%%%%%%%%%%%%%%%%%%%%%%%%%%%%%%%%%%%%%%%%%%
\section{Late-time expansion}\label{sec:expansion}
%%%%%%%%%%%%%%%%%%%%%%%%%%%%%%%%%%%%%%%%%%%%%%%%%%%%%%%%%%%%%%%%%%%%%%%%%%

Late-time evolution of the bubble walls was studied in~\cite{Espinosa:2010hh} using a hydrodynamical approximation in which it is assumed that the cosmic plasma can be modelled as the relativistic perfect fluid, therefore its energy-momentum tensor is given by
\begin{equation}
    T^{\mu \nu}_{\textrm{fluid}} = w u^{\mu}u^{\nu} - g^{\mu\nu}p, \label{eq:energy-momentum_tensor_fluid}
\end{equation}
where $u^{\mu}$ is the four-velocity of the plasma, while $p$ and $w$ are respectively the pressure and the enthalpy. Conservation of energy-momentum projected onto direction $u^{\mu}$ along the flow and the perpendicular one $\bar{u}^{\mu} = \gamma(v,\textbf{v}/v)$ gives the following equations
\begin{align}
    \partial_{\mu}( u^{\mu} w) - u_{\mu}\partial^{\mu}p &= 0,
    \label{eq:bag1}\\
    \bar{u}^{\nu}u^{\mu} w\partial_{\mu}u_{\nu} - \bar{u}^{\nu}\partial_{\mu}p &= 0.
    \label{eq:bag2}
\end{align}
with $\bar{u}_{\mu}u^{\mu} = 0$ and $\bar{u}^2 = -1$. As the steady-state profile has no characteristic length scale, the solution should depend only on the self-similar variable $\xi = r/t$, where $r$ denotes the distance from the centre of the bubble and $t$ is the time since nucleation. Variable $\xi$ can be interpreted as the velocity of a given point in the profile, while the plasma at the point described by $\xi$ moves with velocity $v(\xi)$. Under this assumptions, equations \eqref{eq:bag1} and \eqref{eq:bag2} can be rewritten as
\begin{align}
    (\xi-v)\frac{\partial_{\xi} e}{w} &= 2\frac{v}{\xi}+[1-\gamma^2v(\xi-v)]\partial_{\xi}v,\\
    (1-v\xi)\frac{\partial_{\xi} p}{w} & = \gamma^2(\xi-v)\partial_{\xi}v
\end{align}
and using the definition of the speed of sound in the plasma $c_s \equiv \frac{\textrm{d}p}{dT}/\frac{\textrm{d}e}{dT}$, they can be combined into the single hydrodynamic equation describing plasma velocity profile $v(\xi)$ in the frame of the bubble centre
\begin{equation} 
2\frac{v}{\xi} =\gamma^2 (1 - v \xi )\left[\frac{\mu^2}{c_s^2}-1\right]\partial_{\xi}v,
\label{eq:bag_central}
\end{equation}
with $\mu = \frac{\xi-v}{1-\xi v}$ denoting the Lorentz-transformed fluid velocity. To proceed further, one has to define the equation of state for the plasma. The most popular choice is the so-called bag model \cite{Espinosa:2010hh, Konstandin:2010dm}, where the speed of sound in both phases is constant and equal $c_s^2 = 1/3$, however, more complex options were recently considered \cite{Ai:2021kak, Ai:2023see}.

\begin{figure}
    \begin{minipage}{0.3\textwidth}
        \centering
        \includegraphics[scale=0.75]{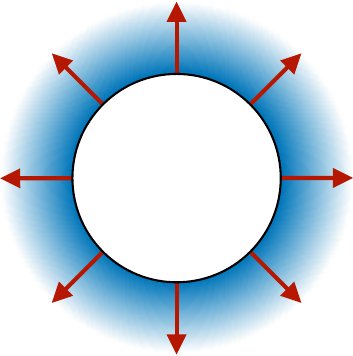}
        \textrm{Deflagration: $v_w<c_s$}
        \label{subfig:def}
    \end{minipage}
    \hfill
    \begin{minipage}{0.3\textwidth}
        \centering
        \captionsetup{labelformat=empty, labelsep = none}
        \includegraphics[scale=0.75]{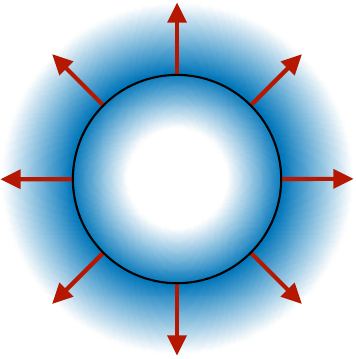}
        \textrm{Hybrid: $c_s<v_w<c_J$}
        \label{subfig:hyb}
    \end{minipage}
    \hfill
    \begin{minipage}{0.3\textwidth}
        \centering
        \includegraphics[scale=0.75]{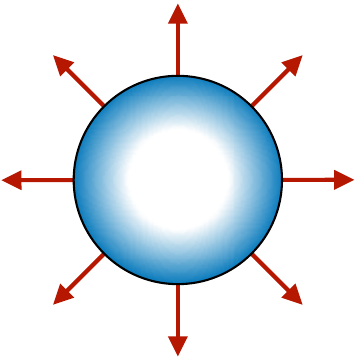}
         \textrm{Detonation: $c_J<v_w$}
        \label{subfig:det}
    \end{minipage}
    \caption{\textit{Schematic representation of three different types of expanding bubbles. Colour saturation denotes the value of the plasma velocity, while black circles represent the position of the bubble wall.}}
    \label{fig:schemes}
\end{figure}

Solutions of equation \eqref{eq:bag_central} with the bag-model equation of state depend only on the transition strength $\alpha$ and bubble-wall velocity in the stationary state $\xi_w$. Possible types of solutions are subsonic deflagrations or supersonic detonations and hybrids schematically depicted in Fig. \ref{fig:schemes}. The velocity at which the shell around the bubble disappears and the solution shifts from hybrid to detonation is given by the Chapman-Jouguet value~\cite{Ai:2023see}:
\begin{equation}\label{eq:vJ}
    c_J = \frac{1+\sqrt{3\alpha_{\bar{\theta}}( 1 - c_s^2 + c_s^2\alpha_{\bar{\theta}}) }}{1/c_s+3c_s\alpha_{\bar{\theta}}}.
\end{equation}
For a detailed discussion of bag model solutions see \cite{Espinosa:2010hh}.

In order to estimate bubble-wall velocity, it is necessary to estimate the total force exerted on a bubble wall from the plasma surrounding the bubble and compare it with a driving force coming from the pressure difference between the false and true vacuum. The equation of motion of a scalar field reads
\begin{equation}
       \Box \phi + \frac{\partial V_{\textrm{eff}}}{\partial \phi} + \sum_{i}\frac{\textrm{d} m_i^2(\phi)}{\textrm{d}\phi}\int\frac{\textrm{d}^3 p}{(2\pi)^3 2 E_i}\delta f_i(p,x) = 0,
       \label{eq:EoM-general}
    \end{equation}
where the index $i$ runs over all the species in the plasma and $\delta f_i$ is the non-equilibrium part of the particle distribution function as the equilibrium part is already included in the effective potential \cite{Ai:2021kak}. Assuming a planar wall expanding with a constant velocity in $z$ direction, the equation above can be written as
\begin{equation}
        \int \textrm{d}z\frac{\textrm{d}\phi}{\textrm{d}z} \left(\Box \phi + \frac{\partial V_{\textrm{eff}}}{\partial \phi} + \sum_{i}\frac{\textrm{d} m_i^2(\phi)}{\textrm{d}\phi}\int\frac{\textrm{d}^3 p}{(2\pi)^3 2 E_i}\delta f_i(p,x)\right) = 0
    \end{equation}
and integrated leading to 
\begin{equation}
\label{eq:backreaction_force}
        \Delta V_{\textrm{eff}} = \int \textrm{d} z \frac{\partial V_{\textrm{eff}}}{\partial T}\frac{\textrm{d}T}{\textrm{d}z} - \sum_i \int \textrm{d} \phi \frac{\textrm{d}m_i^2(\phi)}{\textrm{d}\phi}\int\frac{\textrm{d}^3 p}{(2\pi)^3 2 E_i}\delta f_i(p,x).
    \end{equation}
The left-hand side can be interpreted as the pressure difference between the true vacuum inside the bubble and the false vacuum outside. The right-hand side is the backreaction force and can be separated into the equilibrium part coming from heated plasma at the bubble front (first term) and typically subdominant~\cite{Laurent:2022jrs} non-equilibrium friction (second term).

Steady-state bubble-wall velocity can be determined by solving the equation of motion for the scalar field \eqref{eq:EoM-general} together with the Boltzmann equations for all the particle species in the plasma. This approach is computationally demanding and typically the so-called fluid ansatz is used and solutions are found by looking for the configuration for which equation of motion has two vanishing moments (see \cite{Friedlander:2020tnq, Cline:2021iff, Lewicki:2021pgr, Laurent:2020gpg, Laurent:2022jrs} for more details). Full computations of the bubble-wall velocity including out-of-equilibrium friction are challenging, therefore to simplify the problem, it is often assumed that $\delta f_i = 0$, which is called the local thermal equilibrium (LTE) scenario.

It was recently shown that assuming the LTE approximation and using the conservation of entropy it is possible to determine the bubble wall velocity in a steady state in a much simpler way. The first attempt based on the bag equation of state~\cite{Ai:2021kak}, has been already generalized to a more broad class of the equations of state and beyond the planar wall limit~\cite{Ai:2023see}. It has been also shown, that the bubble-wall velocity can be estimated based on a few parameters evaluated at the nucleation temperature: enthalpy ratio between phases $\psi_N = \frac{\omega_b}{\omega_s}$, transition strength $\alpha_{\bar{\theta}}$ and speed of sound in both phases $c_s$ and $c_b$~\cite{Ai:2023see}.
Nevertheless, both non-equilibrium Boltzmann methods as well as simplified equilibrium approximations are based on the assumption, that stationary hydrodynamical behaviour sets in just after nucleation and do not take into account the early evolution of the system, as all the calculations are performed for stationary profiles. 

\section{Benchmark model: SM + real scalar singlet}\label{sec:singlet}
Probably the simplest and most studied model in which a strong first-order electroweak phase transition can be realized is the real scalar singlet extension of the Standard Model~\cite{McDonald:1993ey, Espinosa:1993bs, Espinosa:2007qk, Profumo:2007wc, Espinosa:2011ax, Barger:2011vm, Cline:2012hg, Alanne:2014bra, Curtin:2014jma, Vaskonen:2016yiu, Kurup:2017dzf, Beniwal:2017eik, Beniwal:2018hyi, Niemi:2021qvp}, where in addition to the Standard
Model Higgs doublet, the scalar sector includes the $Z_2-$symmetric scalar field $s$. The tree-level scalar potential of this theory in a unitary gauge is given by
\begin{equation}
\label{eq:tree-level}
    V_0(h,s) = \frac{1}{2} \mu_h^2 h^2 + \frac{1}{4}\lambda_h h^4 + \frac{1}{4} \lambda_{hs} h^2 s^2 + \frac{1}{2} \mu_s^2 s^2 + \frac{1}{4}\lambda_{s}s^4.
\end{equation}
The mass term for the Higgs $\mu_h$ and the quartic coupling $\lambda_h$ are fixed in such a way, that in the electroweak vacuum is $(h,s)=(\upsilon, 0)$ with $\upsilon = 246.2$ GeV and the physical mass of the Higgs is $m_h = 125.09$ GeV, which leads to
\begin{displaymath}
\lambda_h = \frac{m_h^2}{2v^2} \qquad\textrm{and}\qquad \mu_h^2 = -\lambda_h\upsilon^2,
\end{displaymath}
leaving scalar singlet mass $m_s$, its quartic coupling $\lambda_s$ and the portal coupling between the Higgs and the scalar singlet $\lambda_{hs}$ as three free parameters of the model.
To study the phase transition, thermal corrections to the potential need to be included:
\begin{equation}
    V_{\textrm{eff}}(h,s,T) = V_0(h,s) + \sum_i \frac{n_i T^4}{2\pi^2}J_{b/f}\left(\frac{m_i(h,s)}{T}\right), 
\end{equation}
where the sum runs over all particle species of the model, and for all species $n_i$ is the number of degrees of freedom, $m_i(h,s)$ is its field-dependent mass, and $J_{b/f}$ is the thermal function given by 
\begin{equation}
    J_{b/f}(x) = \pm \int_{0}^{\infty}\textrm{d}y y^2\log\left( 1 \mp \exp\left( - \sqrt{y^2 + x^2}\right) \right)\, ,
\end{equation}
where the upper (lower) sign is for bosons (fermions). 
Expanding the thermal function in the relativistic regime ($T\gg m$) gives
\begin{align}
\label{eq:J_high}
    J_b(x)\approx & -\frac{\pi^4}{45} + \frac{\pi^2}{12} x^2 +\mathcal{O}(x^3) &
    J_f(x)\approx & -\frac{7}{8} \frac{\pi^4}{45} + \frac{\pi^2}{24} x^2+\mathcal{O}(x^4\log x^2)\, .
\end{align}
Therefore, the high temperature expansion \eqref{eq:J_high} yields the well-known result 
\begin{equation}
    V_{\textrm{eff}}(h,s,T) \approx V_0(h,s) -\frac{g_{\ast}\pi^2}{90}T^4 + \sum_i \frac{c_i n_i}{24} m_i^2(h,s) T^2
\end{equation}
with $c_i = 1/2$ for fermions, $c_i = 1$ for bosons and $g_{\ast}$ denoting the effective number of relativistic degrees of freedom, namely
\begin{equation}
g_{\ast}(T) = \sum_{i\in\textrm{bosons}}g_i\left(\frac{T_i}{T}\right)^4+ \sum_{i\in\textrm{fermions}}\frac{7}{8}g_i\left(\frac{T_i}{T}\right)^4.
\end{equation} 
Finally, the whole effective potential of the model can be written in compact form as the tree-level potential \eqref{eq:tree-level} with temperature-dependent mass terms
\begin{equation}
    \mu_h^2(T) \coloneqq \mu_h^2 + c_h^2 T^2 \qquad \textrm{and} \qquad \mu_s^2(T) \coloneqq \mu_s^2 + c_s^2 T^2,
\end{equation}
with
\begin{equation}
    c_h^2 = \frac{1}{48}\left(9g^2 + 3g'^2 + 12y_t^2 + 24\lambda_{h} + 2 \lambda_{hs}\right) \qquad \textrm{and} \qquad c_s^2 = \frac{1}{12} \left( 2 \lambda_{hs} + 3\lambda_s \right),  
\end{equation}
where $g$ and $g'$ are electroweak couplings and $y_t$ is the Yukawa coupling for the top quark. This leads to the high-temperature approximation of the effective potential in a compact form
\begin{equation}
\begin{aligned}
\label{eq:potential}
    V_{\textrm{eff}}(h,s,T) =& -\frac{g_{\ast}\pi^2}{90}T^4 + \frac{1}{2} (\mu_h^2 + c_h^2 T^2) h^2 + \frac{1}{4}\lambda_h h^4 + \frac{1}{4} \lambda_{hs} h^2 s^2 +\\ &+ \frac{1}{2} (\mu_s^2 + c_s^2 T^2) s^2 + \frac{1}{4}\lambda_{s}s^4\, .
\end{aligned}
\end{equation}
In this simplified treatment we neglect the presence of the Coleman-Weinberg corrections to the effective potential.

%%%%%%%%%%%%%%%%%%%%%%%%%%%%%%%%%%%%%%%%%%%%
\begin{wrapfigure}{l}{0.5\textwidth}
\centering
\includegraphics{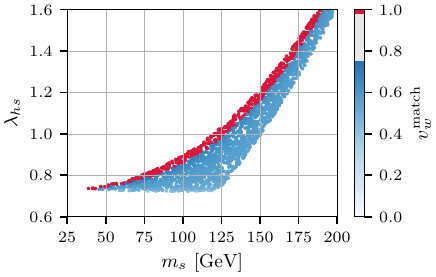}
\caption{\textit{Bubble-wall velocities for the scalar singlet model in local thermal equilibrium determined using the matching method derived in~\cite{Ai:2023see} which assumes a steady state solution at all times. Deflagrations and hybrids are represented by blue dots with the shade depicting the wall velocity, while runaways are marked with red. No stationary detonations were found which gives rise to the gray gap in velocities.}}
\label{fig:analytical_velocities}
\vspace{-20pt}
\end{wrapfigure}
%%%%%%%%%%%%%%%%%%%%%%%%%%%%%%%%%%%%%%%%%%%%%%%%%

To prepare a sample for our study we performed a scan over the scalar singlet mass $m_s$ and portal coupling $\lambda_{hs}$ fixing the quartic coupling to $\lambda_s = 1$ with the use of the \verb+CosmoTransitions+ package \cite{Wainwright:2011kj}. We determine the critical and nucleation temperatures as well as the transition strength for each point.

% \begin{figure}[t]
%     \centering
%     \includegraphics{Graphics/scan_velocity.pdf}
%     \caption{\textit{Bubble-wall velocities for the scalar singlet model in local thermal equilibrium determined using the matching method derived in~\cite{Ai:2023see} which assumes a steady state solution at all times. Deflagrations and hybrids are represented by blue dots with the shade depicting the wall velocity, while runaways are marked with red. No stationary detonations were found which gives rise to the gray gap in velocities.}}
%     \label{fig:analytical_velocities}
% \end{figure}

Next, using approximation derived in~\cite{Ai:2023see}, we determine the bubble-wall velocities under the assumption of preserving local thermal equilibrium, finding a variety of deflagration and hybrid solutions, as well as runaway scenarios for the strongest transitions from the investigated sample. The results are presented in Fig. \ref{fig:analytical_velocities}, where the shades of blue indicate the wall velocity while the red points mark the runaway solutions.

\section{Real-time dynamics of the system}\label{sec:dynamics}
In hydrodynamical treatment, we describe the plasma as a perfect fluid with temperature $T$, characterized by the internal energy density $e$, pressure $p$ and enthalpy density $w$. As the effective potential $V_{\textrm{eff}}(h, s, T)$ can be interpreted as the free energy density $\mathcal{F}$ of the system, we can define\footnote{Note that in contrast to the previous papers, the contributions from relativistic degrees of freedom are already included in the definition of the effective potential \eqref{eq:potential} and do not need to be added separately.}
\begin{align}
p(h, s, T) &= - V_{\textrm{eff}}(h, s,T),\label{eq:pressure}\\
e(h, s, T) &= V_{\textrm{eff}}(h, s, T) - T\frac{\textrm{d}V_{\textrm{eff}}(h, s, T)}{\textrm{d}T}, \label{eq:energy}\\
w(h, s, T) &= -T\frac{\textrm{d}V_{\textrm{eff}}(h, s, T)}{\textrm{d}T}\label{eq:entalpy}.
\end{align}

The total energy-momentum tensor of the system is a sum of energy-momentum tensors for the fields~\footnote{Note that in our convention the vacuum energy is a part of the fluid energy-momentum tensor~\eqref{eq:energy-momentum_tensor_fluid} and is not present here.}:
%\begin{align}
\begin{equation}
    T^{\mu \nu}_{\textrm{fields}} = \partial^\mu h \partial^\nu h + \partial^\mu s \partial^\nu s - g^{\mu \nu}\left( \frac{1}{2} \partial_\alpha h \partial^\alpha h + \frac{1}{2} \partial_\alpha s \partial^\alpha s \right)\, , %\\
%    T^{\mu \nu}_{\textrm{fluid}} &= w u^\mu u^\nu + g^{\mu \nu} p\, ,
\end{equation}
%\end{align}
and the one of the fluid given by \eqref{eq:energy-momentum_tensor_fluid}.

The energy-momentum tensor of the system is conserved $(\nabla_\mu T^{\mu \nu} = 0 )$, however, both contributions are not conserved separately:
\begin{equation}
    \nabla_\mu T^{\mu \nu}_{\textrm{fields}} = \frac{\partial V_{\textrm{eff}}}{\partial h} \partial^\nu h +  \frac{\partial V_{\textrm{eff}}}{\partial s} \partial^\nu s = - \nabla_\mu T^{\mu \nu}_{\textrm{fluid}}. \label{eq:energy_momentum_conservation}
\end{equation}
Note that as we are interested in the local thermal equilibrium scenario, here the energy transfer between the scalar field and the plasma is possible only through the temperature-dependent effective potential, and there is no additional effective friction typically added to capture non-equilibrium effects \cite{Krajewski:2023clt}.
The left equality of \eqref{eq:energy_momentum_conservation} is satisfied if fields $h$ and $s$ follow standard wave equations which in spherical coordinates take the form
\begin{subequations}
\label{eq:fields}
\begin{align}
    &\begin{aligned}
          -\partial_t^2 h + \frac{1}{r^2}\partial_r(r^2\partial_r h) -
\frac{\partial V_{\textrm{eff}}}{\partial h} = 0, 
    &\end{aligned}\\
    &\begin{aligned}
          -\partial_t^2 s + \frac{1}{r^2}\partial_r(r^2\partial_r s) -
\frac{\partial V_{\textrm{eff}}}{\partial s} = 0.
    &\end{aligned}\label{eq:nonlinear_wave_equation}
\end{align}
\end{subequations}
Due to the spherical symmetry of our problem, we assume that the four-velocity of the perfect fluid is of the form $ u = (\gamma, \gamma v, 0, 0)^T$ with $\gamma:= (1 - v^2)^{-1/2}$. We will determine the equations governing the evolution of two parameters $v$ and $p$ considering temporal ($\nu = 0$) and radial component ($\nu=1$) of Eq.~\eqref{eq:energy_momentum_conservation}.
Introducing new variables $Z:=w\gamma^2v$ and $\tau:=w\gamma^2 - p$ we get
\begin{subequations}\label{eq:plasma_equations_of_motion}
\begin{align}
&\begin{aligned}
\nabla_\mu T^{\mu 0}_{\textrm{fluid}} &= \partial_t \tau + \frac{1}{r^2} \partial_r (r^2 (\tau + p) v)= \frac{\partial V_{\textrm{eff}}}{\partial h} \partial_t h +  \frac{\partial V_{\textrm{eff}}}{\partial s} \partial_t s,\\
\end{aligned}\\
&\begin{aligned}
\nabla_\mu T^{\mu 1}_{\textrm{fluid}} &= \partial_t Z + \frac{1}{r^2} \partial_r \left(r^2 Zv \right) + \partial_r p= -\frac{\partial V_{\textrm{eff}}}{\partial h} \partial_r h -\frac{\partial V_{\textrm{eff}}}{\partial s} \partial_r s.
\end{aligned}
\end{align}
\end{subequations}
The system of equations for the fields \eqref{eq:fields} and for the plasma \eqref{eq:plasma_equations_of_motion}, together with the equations of state \eqref{eq:pressure} and \eqref{eq:energy} define the dynamics of the growing bubble and was solved numerically on the lattice. For details of numerical treatment, see Appendix \ref{appendix}.  

\section{Results of the simulations}\label{sec:simulations}
We use our code~\cite{Krajewski:2023clt} to perform real-time simulations of the bubble-wall expansion for all the points shown in Fig.~\ref{fig:analytical_velocities}. Each simulation was initialized with fields profiles corresponding to the nucleated critical bubble, while plasma velocity and temperature were fixed as spatially uniform, $v(r)=0$ and $T(r) = T_n$ respectively. This gives a good approximation of homogeneous plasma in which nucleation takes place and allows us to carefully follow the early stages of the evolution of the bubbles and formation of the fluid profiles. %\footnote{Ref.~\cite{Balaji:2020yrx} used a similar setup and solved the equations of motion for the very early stages of the evolution, showing that subsonic plasma profiles may start to form.}. 
This is not possible with different methods of determining bubble-wall velocities, which look for the stationary states only.

\begin{figure}[t]
    \centering
    \includegraphics{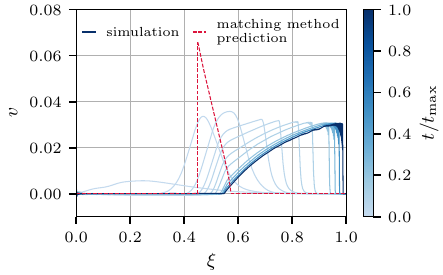}
    \includegraphics{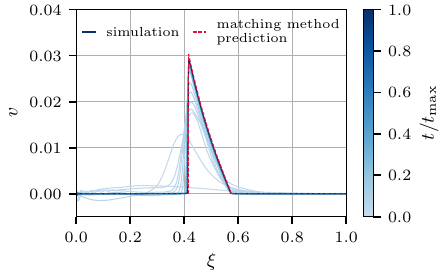}
    \caption{\textit{Two possible scenarios for the growing bubble in local thermal equilibrium approximation: rapid expansion beyond Chapman-Jouguet velocity leading to a runaway scenario (left panel) and evolution toward a stationary state predicted by matching conditions (right panel). Solid blue curves represent the results of the real-time simulations, where darker shades correspond to later times. Red, dashed curves denote predictions of the equilibrium methods.}}
    \label{fig:evolution}
\end{figure}

\begin{figure}[t]
    \centering
    \includegraphics[scale=0.98]{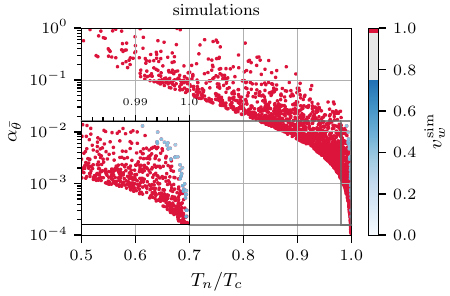}
    \includegraphics[scale=0.98]{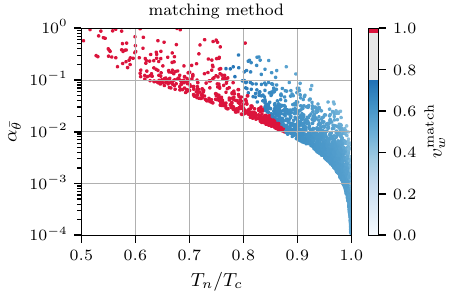}
    \caption{\textit{Results for the bubble wall velocity in LTE from the hydrodynamical simulations (left panel) and from analytical calculations~\cite{Ai:2023see} (right panel). The stationary states are shown in shades of blue corresponding to the wall velocity in both panels while runaway scenarios are shown in red. In most of the parameter space where matching equations predict a deflagration or hybrid solution the simulation results in a runaway. In those cases, the simulated bubble accelerates beyond the Jouguet velocity before the heated fluid shell around the bubble is formed while analytical methods assume a steady state at all times where the shell is heated enough to cease the acceleration.}}
    \label{fig:stationary_states}
\end{figure}

Fig.~\ref{fig:stationary_states} highlights the difference between our results in the left panel and the matching method of \cite{Ai:2023see} in the right. For majority of points where the matching method predicts deflagrations and hybrids, we observe a very different behaviour. The bubble walls accelerate beyond the Jouguet velocity before the heated fluid shell in front of the bubble is formed. The corresponding evolution for a benchmark is shown in the left panel of the Fig.~\ref{fig:evolution}. This phenomenon is easy to understand as the bubble growth in local thermal equilibrium is very rapid. During the early stages of the evolution, the amplitude of plasma profiles is much smaller, and profiles are less sharp than the ones predicted by the steady-state solutions. This leads to a significantly lower backreaction force (see eq. \eqref{eq:backreaction_force}), which allows the bubble wall to accelerate further. If the profile accelerates above the Jouguet velocity before the steady state profile is reached, the stationary state will not be achieved and the bubble will continue to grow in the runaway regime. This occurs for the vast majority of the points from the scan where the matching method can be used to compute the wall velocity.

%%%%%%%%%%%%%%%%%%%%%%%%%%%%%%%%%%%%%%%%%%%%
\begin{wrapfigure}{l}{0.5\textwidth}
\vspace{-10pt}
\centering
\includegraphics{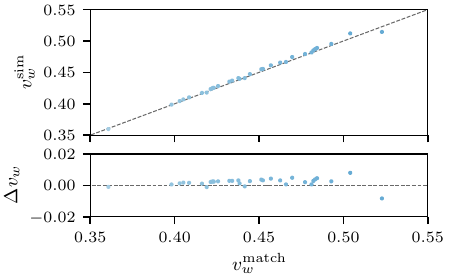}
\caption{\textit{ Comparison of the bubble wall velocity from matching method estimates $v^{\rm match}_w$ \cite{Ai:2023see} vs those measured in hydrodynamic simulation $v_w^{\textrm{sim}}$. The lower panel shows the difference between these two values $\Delta v_w = v_w^{\textrm{sim}} - v_w^{\textrm{match}} $. The low number of points results from the fact that the two methods agree on the final steady state solution in a very small part of the parameter space where $T_n/T_c$ and $\alpha_{\bar{\theta}}$ are tuned. The result of the simulation is much more often a runaway (see Fig.~\ref{fig:stationary_states}).}}
\label{fig:velocity_comparison}
\vspace{-10pt}
\end{wrapfigure}
%%%%%%%%%%%%%%%%%%%%%%%%%%%%%%%%%%%%%%%%%%%%%%%%%

For moderately strong transitions which at the same time are not supercooled, the temperature of the plasma around
the bubble front reaches more quickly the critical temperature $T_c$. This stops further acceleration of the bubble wall, as larger wall velocity (up to Chapman-Jouguet velocity) would demand higher temperature in the peak exceeding $T_c$ which would locally tend to reverse the ongoing transition.
This mechanism is called the hydrodynamical obstruction~\cite{Konstandin:2010dm} and has already been demonstrated in hydrodynamical simulations~\cite{Krajewski:2023clt} leading to velocity gaps within possible expansion modes. From this moment the bubble starts to asymptote to a constant velocity and the plasma profiles evolve towards stationary states predicted by the matching conditions. 
% \begin{figure}[t]
%     \centering
%     \includegraphics[scale=0.98]{Graphics/velocity_comparison_all.pdf}
%     \caption{\textit{ Comparison of the bubble wall velocity from matching method estimates $v^{\rm match}_w$ \cite{Ai:2023see} vs those measured in hydrodynamic simulation $v_w^{\textrm{sim}}$. The lower panel shows the difference between these two values $\Delta v_w = v_w^{\textrm{sim}} - v_w^{\textrm{match}} $. The low number of points results from the fact that the two methods agree on the final steady state solution in a very small part of the parameter space where $T_n/T_c$ and $\alpha_{\bar{\theta}}$ are tuned. The result of the simulation is much more often a runaway (see Fig.~\ref{fig:stationary_states}).}}
%     \label{fig:velocity_comparison}
% \end{figure}
The evolution of the plasma velocity profile towards a stationary state compared with the analytically predicted profile for a benchmark transition is shown in the right panel of Fig.~\ref{fig:evolution}. While this is the case only in a very small part of the parameter space, it is worth noting that bubble-wall velocities as well as plasma profiles obtained in the simulations are in very good agreement with those found with the matching method of~\cite{Ai:2023see} wherever they both find non-runaway behaviour. We show a direct comparison of the predictions for this set of points in Fig.~\ref{fig:velocity_comparison} finding the difference to be of the order of a few per cent at most.

Ref.~\cite{Balaji:2020yrx} used a similar setup and solved the equations of motion for the very early stages of evolution ($t_{max} \approx 0.6$ GeV$^{-1}$, compared with $t_{max} \simeq 100$ GeV$^{-1}$ in this work.). It was shown the profiles began to form, however, the dynamical range was not large enough to verify if they reached steady-state solutions. It is also important to point out that the authors %of \cite{Balaji:2020yrx},
focused on a single benchmark with $T_n/T_c \approx 0.994$ and observed the impact of hydrodynamical obstruction, which is consistent with our results. Ref.~\cite{Jinno:2022mie} also observed relaxation of the fluid to form a shell, although, in that work, the code used a fixed value for the wall velocity and the formation of a steady-state profile was only the result of this prior assumption. 

\section{Conclusions}\label{sec:conclusions}

We have studied the growth of cosmological bubbles nucleated during first-order phase transitions using real-time hydrodynamic simulations. Focusing on the bubble-wall velocity in the local thermal equilibrium, we have confirmed that pure equilibrium backreaction can lead to a steady-state solution. If this is the case and the stationary state is realised in our simulation, the final velocity agrees very well with recent predictions based on the matching equations~\cite{Ai:2023see}. Such scenarios are however very rare and demand fine-tuning of nucleation temperature as the stationary expansion is the outcome only in the absence of any supercooling ($T_n/T_c \approx 1$). %It is also important to point out that the authors of \cite{Balaji:2020yrx}, focused on a single benchmark with $T_n/T_c \approx 0.994$ and observed similar impact of hydrodynamical obstruction, which is consistent with our results discussed in this paper.

We have found that generically without a non-equilibrium friction bubbles expand as runaways. In the very early stages of their evolution, the nucleated bubbles accelerate quickly as the heated fluid shell around them is just beginning to form. Typically the walls accelerate past the Jouguet velocity and start to develop into detonations before the fluid around them forms into a familiar steady state profile. Once this occurs the bubble will behave as a runaway despite the fact that a solution exists for which the steady-state profile would match the vacuum pressure and cease the acceleration at a lower velocity.  

We have focused on a very simple extension of the Standard Model with a neutral singlet to illustrate the prevalence of the early runaway scenario. We have found that nearly the entire parameter space of the model predicting a first-order transition would result in a runaway if one neglects the non-equilibrium friction. The only exceptions are tuned scenarios where the transition is relatively strong yet nucleation occurs extremely close to the critical temperature. 

We have shown that the early stages of the evolution of bubbles in cosmological phase transitions can have a crucial impact on the resulting phenomenology. In particular, assuming a local thermal equilibrium at those early times generically leads to a runaway solution even in cases where analytical methods using fully developed steady-state profiles suggest a small terminal velocity. This would have serious consequences for the predictions of models concerning the possible production of baryon asymmetry in electroweak baryogenesis, as well as on the gravitational wave signals generated in first-order phase transitions.

\subsubsection*{Acknowledgements}
M.Z. would like to thank the organisers of the workshop "How fast does the bubble grow?" at DESY, Hamburg for their hospitality. This work was supported by the Polish National Agency for Academic Exchange within Polish Returns Programme under agreement PPN/PPO/2020/1/00013/U/00001 and the Polish National Science Center grant 2018/31/D/ST2/02048. T.K. was supported by grant 2019/33/B/ST9/01564 from the Polish National Science Centre.

\appendix
\section{Numerical treatment}\label{appendix}
Equations \eqref{eq:fields} and \eqref{eq:plasma_equations_of_motion} describing the dynamics of the model under consideration are highly nonlinear and the exact solution of the system is not known. In order to get a detailed understanding of the evolution of bubbles described by the set of equations we solve the system numerically extending the code used in our previous studies~\cite{Krajewski:2023clt} to the case of two scalar fields. The extension is straight forward and to keep this manuscript self-contained we will now only briefly describe used methods.

Equations \eqref{eq:fields} and \eqref{eq:plasma_equations_of_motion} have to be supplemented by the proper boundary conditions. We chose Neumann boundary conditions for both scalar fields at the centre of the bubble which correspond to $r=0$. In general, the Dirichlet boundary conditions corresponding to the false vacuum expectation values should be imposed at an infinite distance $r = \infty$ from the nucleation site. In order to treat the problem numerically, we had to limit our computational domain to be a finite interval $[0, R]$ with $R$ large enough to be outside the light cone of the nucleated bubble, so imposing boundary conditions at $r=R$ is physically equivalent to $r=\infty$ for a finite period of time simulated in our code.

The initial conditions for our simulations are critical bubble profiles calculated using \verb+CosmoTransitions+ code \cite{Wainwright:2011kj}. The critical bubble which is at the boundary between collapsing and expanding bubbles is an unstable static solution of equations of motion. Due to numerical imperfections, the approximate profile may collapse instead of growing and a minimal stretching of the initial profile is necessary to guarantee the expansion of the bubble. The conserved variables $Z$ and $\tau$ describing the state of the plasma are initialized to the values corresponding to the plasma with temperature $T$ equal to nucleation temperature, staying at rest $v=0$ with respect to the nucleation site.

We used the Galerkin discontinuous method to discretize  equations~\eqref{eq:fields} and~\eqref{eq:plasma_equations_of_motion} in space. We used the so-called mixed formulation of the equations of motion for the fields~\eqref{eq:fields} (which are nonlinear wave equations), so the gradients of the fields and their values are discretized as independent variables. Both field values and quantities $Z$ and $\tau$ describing plasma are interpolated using piecewise constant functions. Piecewise linear functions are used for gradients of fields. The integrals in weak forms of discretized equations are approximated by numerical quadratures which are exact for elements of the base of space of interpolation functions. The numerical fluxes were chosen such that the obtained method is the generalization of the central finite difference scheme (for planar bubble walls described in Cartesian coordinates the method would recover the central finite difference scheme of second order).

\begin{figure}[t]
    \centering
    \includegraphics[scale = 0.6]{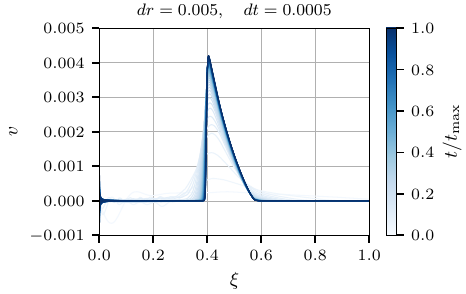}
    \includegraphics[scale = 0.6]{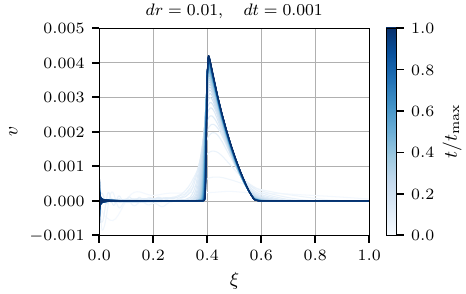}
    \includegraphics[scale = 0.6]{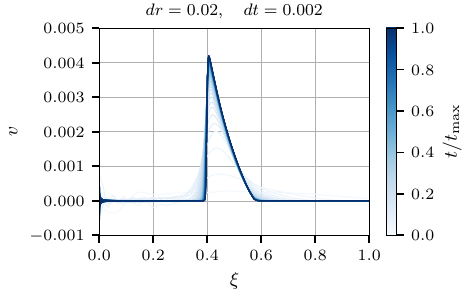}
    \includegraphics[scale = 0.6]{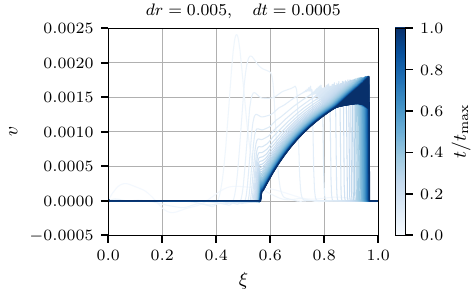}
    \includegraphics[scale = 0.6]{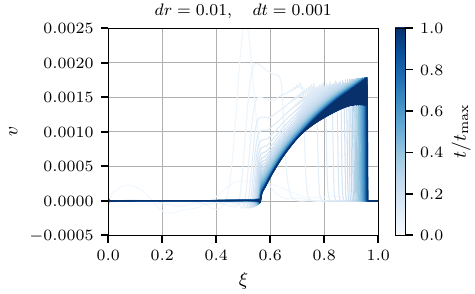}
    \includegraphics[scale = 0.6]{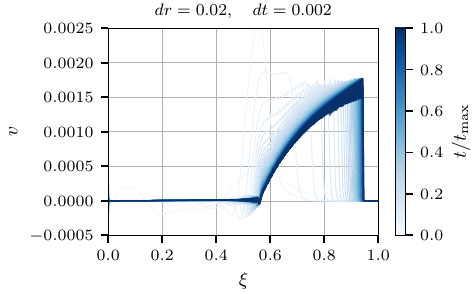}
    \caption{\textit{Convergence analysis: velocity profile evolution on different lattices for typical deflagration (upper row) and runaway (lower row).  Three different grids are compared: fine grid, the default one used in this work and thick grid. Shape of the profile does not depend on the choice of the grid. For ultra-relativistic bubble-wall, numerical oscillations coming from the Gibbs effect are more visible, however they do not impact the overall result.}}
    \label{fig:convergence}
\end{figure}

Wave equations \eqref{eq:fields} describing the evolution of scalar fields were discretized using a well-known position version of St\"omer-Verlet scheme. It can be formulated as the right-discontinuous Galerkin method in time. This formulation was used in our previous paper~\cite{Krajewski:2023clt} to derive in a self-consistent manner the discretization of effective friction terms which, however, are not used in the current study (they violate the LTE assumption).

The discretization in time of \eqref{eq:plasma_equations_of_motion} is more involved, and no extension with respect to the code used in Ref~\cite{Krajewski:2023clt} needs to be done. Equations~\eqref{eq:plasma_equations_of_motion} are first-order hyperbolic equations which are challenging for numerical methods due to the formation of discontinuities (shocks, etc.) in a~finite time period of time from smooth initial conditions. Many various numerical methods dealing with this problem were proposed in the past. We chose to use flux-corrected transport to keep second-order precision in smooth regions of the profile and avoid numerical artefacts (Gibbs effect) around discontinuities. In this approach, we have two schemes, called high and low order, merged in such a way as to keep the solution in the range of physically correct values according to maximum principle \cite{Boris:1973,Zalesak:1979}. The second-order scheme (the high-order one) was integrated in time using the midpoint rule (which is by itself second order). The low order scheme was obtained by algebraic upwinding \cite{Kuzmin:2002, Kuzmin:2003, Moller:2013, Kuzmin:2021} which produces the local extrema diminishing scheme, and was integrated in time using backward (implicit) Euler method. Differences in fluxes computed in the two schemes are called antidiffusive fluxes and are used (after proper limiting) to correct the low-order scheme result. The aim of the limiting procedure is to suppress antidiffusive fluxes in regions where high-order scheme breaks down due to discontinuities. We used Zalesak's peak preserving limiter \cite{Zalesak:2012} corrected by the idea inspired by \cite{Kunhardt:1987} to restrict distances from which the conserved quantity values should be considered to limit antidiffusive fluxes. Details of the construction of the limiter can be found in \cite{Krajewski:2023clt}.

Finally, the conversion from conserved variables $\tau$, $Z$ to primitive ones $T$, $v$ was done analogously as described in \cite{Krajewski:2023clt}. Equations of state \eqref{eq:pressure} and \eqref{eq:entalpy} can be combined to form
\begin{equation}
    \tau + p(h, s, T) - \frac{1}{2}\left(w(h, s, T) +\sqrt{w(h, s, T)^2+ 4Z^2} \right) = 0. \label{eq:conserved_to_primitive}
\end{equation}
For given values of $\tau$, $Z$ (and $h$, $s$) equation \eqref{eq:conserved_to_primitive} can be numerically solved (we used Raphson--Newton method) with respect to $T$. The velocity of the plasma can be recovered from the definition of $Z =  w \gamma^2 v$ and \eqref{eq:entalpy} for a given temperature $T$.

We demonstrate the convergence of our numerical methods on two benchmarks from our scans in Fig~\ref{fig:convergence}. The columns correspond to finer lattice spacings which we see have no impact on the results. The upper row shows a point with $T_n/T_c\approx 1$ in which the wall reaches a finite velocity and a steady state profile forms. The lower row represents a point where the wall passes the Jouguet velocity in its early evolution and runs away.

\bibliographystyle{JHEP}
\bibliography{main} 
\end{document}